\documentclass[12pt]{article}

\usepackage{a4wide}
\usepackage[pdftex,usenames,dvipsnames]{color}
\usepackage{graphics}
\usepackage{amsfonts}
\usepackage{amssymb}

\newcommand{\nit}{\noindent}

\newcommand{\np}{\newpage}
\newcommand{\dsp}{\displaystyle}
\newcommand{\vs}[1]{\vspace{#1 ex}}
\newcommand{\hs}[1]{\hspace{#1 em}}
\newcommand{\bfr}{\begin{flushright}}
\newcommand{\efr}{\end{flushright}}
\newcommand{\bc}{\begin{center}}
\newcommand{\ec}{\end{center}}
\newcommand{\ben}{\begin{enumerate}}
\newcommand{\een}{\end{enumerate}}

\newcommand{\be}{\begin{equation}}
\newcommand{\ee}{\end{equation}}
\newcommand{\ba}{\begin{array}}
\newcommand{\ea}{\end{array}}
\newcommand{\ct}{\cite}

\newcommand{\dd}[2]{\frac{\partial{#1}}{\partial{#2}}}
\newcommand{\ag}{\alpha}

\newcommand{\gam}{\gamma}

\newcommand{\ve}{\varepsilon}

\newcommand{\thg}{\theta}
\newcommand{\kg}{\kappa}

\newcommand{\rg}{\rho}

\newcommand{\vf}{\varphi}

\newcommand{\Og}{\Omega}
\newcommand{\Lb}{\Lambda}

\newcommand{\cA}{{\cal A}}
\newcommand{\cB}{{\cal B}}

\newcommand{\cM}{{\cal M}}

\newcommand{\mfr}{{\sf r}_*}

\newcommand{\lh}{\left(}
\newcommand{\rh}{\right)}
\newcommand{\ld}{\left.}
\newcommand{\rd}{\right.}

\newcommand{\nb}{\nabla}

\begin{document}

\pagestyle{empty}

\bfr
Nikhef/2017-062
\efr
\vs{3}

\bc
{\Large \bf Chaplygin gas halos}
\vs{5}

{\large T.\ de Beer$^{a,b}$ and J.W.\ van Holten$^c$}
\vs{4}

Lorentz Institute \\
Leiden University, Leiden NL

and

Nikhef, Amsterdam NL
\ec
\vs{7}

\nit
{\small 
{\bf Abstract} \\
Unification of dark matter and dark energy as short- and long-range manifestations of a single 
cosmological substance is possible in models described by the generalized Chaplygin gas equation
of state. We show it admits halo-like structures and discuss their density profiles, the resulting 
space-time geometry and the rotational velocity profiles expected in these models.
}

\vfill

\footnoterule
\nit
$^a$ {\em Present address:}  Dept.\ of Physics, Univ.\ of Toronto, 60 St.\ George Str., Toronto, Canada \\
$^b$ {\em e-mail:} tdebeer@physics.utoronto.ca \\
$^c$ {\em e-mail:} vholten@nikhef.nl

\np
~\hfill

\np
\pagestyle{plain} 
\pagenumbering{arabic} 

\nit
{\large \bf 1.\ Introduction}
\vs{1}

\nit
Careful measurements of the observable universe have shown that the list of ingredients contributing 
to the total energy density contains more than radiation, curvature and baryonic/standard-model 
matter \ct{Zwicky1933,Bosma1979,Rubin1980,Clowe2006,Riess1998,Perlmutter1999,Astier2012,
PlanckCollaboration2015}. Assuming the theory of General Relativity to describe space-time 
geometry at astrophysical and cosmological length scales, it appears there are two more ingredients 
which behave qualitatively different, associated with dark matter and dark energy. 

In the $\Lb$CDM-model of cosmology dark matter is commonly associated with a cold gas of 
massive, electrically neutral non-relativistic particles of non-baryonic origin 
\ct{Steigman1985,Peter2012,Ringwald2016}, whilst the dark energy is described by a cosmological 
constant which can be an infrared remnant of unknown fundamental physics 
\ct{Wetterich1988,Wetterich2002,Sahni2002,Verlinde2016,Haridasu2017}. However, although the 
associated length scales and qualitative behavior of dark matter and dark energy are different, there 
is no {\em a priori} reason to include two independent new components. Indeed it is possible to 
construct unified dark matter models (UDM) associating both dark components with a single 
unknown substance. Concrete examples of effective UDM theories are provided by models based 
on the equation of state of the {\em generalized Chaplygin gas} (gCg) \ct{Kamenshchik2001, Bento2002,
Bilic2002,Sandvik2002,alcaniz2003a, alcaniz2003b,Gorini2003,Makler2003,Gorini2008,
Gorini2009,Park2009,El-Zant2015,Bhar2016,Aurich2017,Saha2016,Marttens2017}. 

While it is well-known that the gCg can drive the observed accelerated expansion of the universe, 
see e.g.\ \ct{Makler2003}, to qualify as a dark-matter component as well it should be able to form 
dark-matter like halos in galaxies and galaxy clusters to explain the measured angular velocity 
distribution of stars in galaxies and the non-virial motion of galaxies in clusters. In this paper we 
address the question how to model spherical gCg halos and derive their short- and long-range 
properties to allow comparison with observational constraints on dark matter and dark energy; 
earlier work on these topics can be found in \ct{Gorini2009,El-Zant2015,Bhar2016}.  

This paper is organized as follows. In section 2 we introduce the generalized Chaplygin gas as a 
fluid defined by a specific equation of state. We compute the speed of sound and derive a contraint 
imposed on the parameters in the gCg model by requiring it does not exceed the speed of light. 
We review the cosmological characteristics of a gCg in a homogeneous and isotropic Friedmann 
universe. In section 3 we address the possible existence of spherical halos of a gCg and 
show that their pressure and energy density profiles are governed by a modified form of the 
Tolmann-Oppenheimer-Volkov equation. Expressions for the long- and short-range radial structure 
of such halos are presented and discussed in section 4 and 5. Section 6 describes a modification 
of constraints in the presence of a de Sitter-like horizon to accomodate models in a wider range of 
parameter space; this is followed by a comparison with well-known CMB data in section 7. In 
section 8 we turn to the space-time geometry governed by a gCg-halo structure and derive an 
expression for the rotational velocity profile of test masses on circular orbits. In the final section 9 
we summarize our results and draw conclusions from our analysis. Throughout this paper we use 
natural units in which $c = 1$.

\np
\nit
{\large \bf 2.\ The generalized Chaplygin gas} 
\vs{1}

\nit
The generalized Chaplygin gas is characterized by the equation of state relating pressure $p$ and 
energy density $\ve$ by 
\be
p = - A \ve^{-\ag}, 
\label{2.1}
\ee
where $A$ is a dimensionful proportionality constant and the exponent $\ag$ is a positive number; 
the original Chaplygin gas model was defined with $\ag = 1$ \ct{Chaplygin1904}. The equation of 
state can be converted to a relation between dimensionless quantities by defining a parameter 
$\mu$ with the dimensions of energy density such that $A = \mu^{1 + \ag}$, whence
\be
\frac{p}{\mu} = - \lh \frac{\ve}{\mu} \rh^{-\ag}.
\label{2.2}
\ee
At constant entropy per particle the energy density and pressure are related to the 
number density of particles $\rg$ by relations
\be
\ve = f(\rg), \hs{2} p = \rg f'(\rg) - f(\rg),
\label{2.3}
\ee
such that the pressure is the Legendre transform of the energy density with respect to density $\rg$.
These conditions are solved by 
\be
\frac{\ve}{\mu} = \lh 1 + \lh \frac{\rg}{\rg_0} \rh^{1 + \ag} \rh^{1/(1 + \ag)}, \hs{2}
\frac{p}{\mu} = - \lh 1 + \lh \frac{\rg}{\rg_0} \rh^{1 + \ag} \rh^{-\ag/(1 + \ag)},
\label{2.4}
\ee
where $\rg_0$ is a constant of integration. From these relations one finds the adiabatic speed of 
sound $c_s$ and the equation of state parameter $w$ in the gCg to be given by \ct{Gorini2008}
\be
c_s^2(\rg) = \dd{p}{\ve} = \ag \lh \frac{\ve}{\mu} \rh^{-(1 + \ag)} = - \ag w(\rg),
\label{2.5}
\ee
which is positive for any $\ag > 0$.

In a space-time with metric $g_{\mu\nu}$ the energy-momentum tensor of a perfect fluid obeying 
the gCg equation of state takes the form
\be
T^{\mu\nu} = p g^{\mu\nu} + \lh p + \ve \rh u^{\mu} u^{\nu}, 
\label{2.6}
\ee
where $u^{\mu}$ is the local 4-velocity of the fluid. It follows directly that in the limit of vanishing 
particle density $\rg = 0$ the energy-momentum tensor takes the form of a cosmological constant $\Lb = \mu$:
\be
\ve = - p = \mu \hs{1} \Rightarrow \hs{1} T^{\mu\nu} = - \mu g^{\mu\nu}.
\label{2.7}
\ee
In contrast a dense gCg with $\rg \gg \rg_0$ behaves like a non-relativistic fluid:
\be
\ve \simeq \frac{\mu}{\rg_0}\, \rg, \hs{2} p \simeq 0,
\label{2.8}
\ee
which is the equation of state of a cold gas of non-relativistic particles with mass $m = \mu/\rg_0$. As 
a result the neutral gCg describes a substance which interpolates between dark matter in the dense
early universe and dark energy in the dilute late universe.

This can be seen explicitly by considering a homogeneous gCg in a Friedmann-Lemaitre type universe
with scale factor $a(t)$ and spatial curvature constant $k$:
\be
ds^2 = - dt^2 + a^2 \lh \frac{dr^2}{1 - kr^2} + r^2 d\Og^2 \rh.
\label{2.9}
\ee
In this cosmological setting the covariant conservation of energy-momentum implies
\be
\nb_{\mu} T^{\mu\nu} = 0 \hs{1} \Rightarrow \hs{1} \frac{d (\ve a^3)}{dt} + p \frac{da^3}{dt} = 0.
\label{2.10}
\ee
From this using the gCg equation of state one derives 
\be
\frac{\ve}{\mu} = \left[ 1 + \lh \frac{a_0}{a} \rh^{3(1 + \ag)} \right]^{1/(1 + \ag)},
\label{2.11}
\ee
in agreement with equation (\ref{2.4}) and  ref.\ \ct{Gorini2003}. Clearly, for small $a \ll a_0$ the 
second term in the bracket dominates and $\ve a^3 \simeq$ constant, whilst for large $a \gg a_0$ 
this term is negligible compared to unity and $\ve \simeq \mu$. In fact for $\ag \rightarrow 0$ the 
model reduces to a standard cosmological constant plus a non-relativistic gas like in the 
$\Lb$CDM model:  
\[
\ve \rightarrow \mu + \frac{m}{a^3}, \hs{2} m = \mu a_0^3.
\]
Note that a universal bound $c_s^2 \leq 1$ in eq.\ (\ref{2.5}) implies that, as at late times 
$\ve/\mu \rightarrow 1$:
\[
\ag = c_s^2 \lh \frac{\ve}{\mu} \rh^{1 + \ag} \leq 1. 
\]
\vs{2}

\nit
{\large \bf 3.\ Chaplygin gas halos}
\vs{1} 

\nit
The existence of dark matter is not only suggested by cosmology; in fact the first clear evidence 
came from the average motion of galaxies in clusters \ct{Zwicky1933} and from the motion of 
luminous baryonic matter in the outer regions of spiral galaxies \ct{Bosma1979,Rubin1980}. 
Assuming the mass distribution of galaxies to follow that of luminous matter, the rotation rate of stars 
far from the center of these galaxies violates Kepler's third law. This problem is solved if galaxies 
possess an extended halo of dark matter. Similar amounts of dark matter also explain the apparent 
non-virial motion of galaxies in clusters.

In the context of gCg models this implies that the equation of state (\ref{2.1}) should allow for 
stable non-homogeneous self-gravitating density profiles. In this section we discuss conditions 
for the existence of spherically symmetric self-sustaining halos, neglecting the influence of baryonic 
components. That is, we use the Einstein equations with a source term for a spherical 
non-homogeneous gCg profile to obtain an equation for halo structure. The starting point for our 
discussion is the following spherically symmetric {\em Ansatz} for the space-time metric 
\be
ds^2 = - \cA(r) dt^2 + \cB(r) dr^2 + r^2 d\Og^2.
\label{3.1}
\ee
Taking the energy-momentum tensor of the gCg to be of the form (\ref{2.6}) with non-trivial
radial dependent pressure $p(r)$ and energy density $\ve(r)$, the Einstein equations reduce to
the set
\be
\ba{rl} 
\dsp{ \frac{1}{\cB} \left[ - \frac{1}{r^2} + \frac{\cB}{r^2} + \frac{\cB'}{\cB r} \right] }& 
\hs{-.5} = 8 \pi G \ve, \\
 & \\
\dsp{ \frac{1}{\cB} \left[ \frac{1}{r^2} - \frac{\cB}{r^2} + \frac{\cA'}{\cA r} \right] }& 
\hs{-.5} = 8 \pi G p, \\
 & \\
\dsp{ \frac{1}{2\cB} \left[ \frac{\cA^{\prime\prime}}{\cA} - \frac{\cA'}{2\cA} \lh \frac{\cA'}{\cA} 
 + \frac{\cB'}{\cB} \rh + \frac{1}{r} \lh  \frac{\cA'}{\cA} - \frac{\cB'}{\cB} \rh \right] }&
 \hs{-.5} = 8 \pi G p.
\ea
\label{3.2}
\ee
To solve the first equation we introduce a (non-covariant) mass function defined by 
summing the energy density in excess of $\mu$ up to radius $r$:
\be
\cM(r) = 4\pi \int_0^r d r'\, r^{\prime\, 2} \lh \ve(r') - \mu \rh.
\label{3.3}
\ee
Note that equation (\ref{2.4}) guarantees that the integrand is always non-negative and 
therefore $\cM$ increases monotonically with $r$ from $\cM(0) = 0$. The solution for 
$\cB$ then takes the form 
\be
\cB(r) = \left[ 1 - \frac{2G\cM(r)}{r} - \frac{8\pi G\mu r^2}{3} \right]^{-1}.
\label{3.4}
\ee
This function has a singularity for $r = R$ such that 
\be
 \frac{2G\cM(R)}{R} = 1 - \frac{8\pi G\mu R^2}{3}. 
\label{3.5}
\ee
The singularity at $R$ exists as the right-hand side decreases monotonically as a function 
of $R$ between $R = 0$ and $R = (3/8\pi G \mu)^{1/2}$, whilst on the same interval the 
left-hand side increases semi-monotonically starting from 0 as argued before:
\[
\cM' = 4\pi r^2 \lh \ve(r) - \mu \rh \geq 0, \hs{2} 0 \leq r \leq \sqrt{\frac{3}{8\pi G\mu}}. 
\]
The singularity at $R$ is to be interpreted as cosmic horizon similar to the cosmic horizon
in de Sitter space for an observer located at the origin of co-ordinates. 

To determine $\cM(r)$ and $\cA(r)$ we turn to the other two equations (\ref{3.2}) implying the relations
\be
- \frac{2 p'}{\mu}\, = \frac{(p + \ve)}{\mu}\, \frac{\cA'}{\cA} = 
\frac{2G}{r^2}\,\frac{(p + \ve)}{\mu}\, \frac{\cM + 4 \pi r^3 (p + \mu/3)}{1 - \frac{2G\cM}{r} - \frac{8\pi G\mu r^2}{3}}
\label{3.6}
\ee
This is a modified (reparametrized) form of the Tolman-Oppenheimer-Volkov (TOV) equation applicable 
to the gCg. The original form of the equation was studied extensively in various parameter regimes 
in \ct{Gorini2009}, which also established the existence of a singular radius $R$. In \ct{Bhar2016} 
the original TOV equation was similarly used to search for star-like solutions, which requires different
boundary conditions however. 
\vs{3}

\nit
{\large \bf 4.\ Halo profiles}
\vs{1}

\nit
We now turn to determining the characteristics of the solutions of our modified TOV equation for 
$\cM(r)$, $\ve(r)$ and $p(r)$. We specifically look for radially decreasing solutions of $\ve(r)$ and $p(r)$
which are finite at the horizon $r = R$. The cosmological solution $\rg = 0$ such that $p + \ve = 0$ and 
$p' = 0$ discussed in section 2 trivially satisfies the equation, but does not possess halo structure. 
Due to the non-linear nature of the equation non-trivial exact solutions are hard to find. In developing 
approximations we consider separately the regime of large $r$: $r  \rightarrow R$ near the cosmic 
horizon; and small $r$: $r \rightarrow 0$ near the halo center. In the large-$r$ regime it is convenient 
to introduce a dimensionless parameter $x$: 
\be
r = R(1 - x)
\label{4.1}
\ee
such that $r \rightarrow R$ implies $x \rightarrow 0$. We can then rewrite the TOV equation 
in the form
\be
\ba{l} 
\dsp{ \lh 1 - \frac{2G\cM}{R(1 - x)} - \frac{8\pi G\mu}{3}\, R^2 \lh 1 - x \rh^2 \rh  \frac{d}{dx} \frac{p}{\mu} }\\
 \\
\dsp{ \hs{3} = 4 \pi G \mu R^2 \lh 1 - x \rh \lh \frac{1}{3} + \frac{p}{\mu} + \frac{\cM}{4\pi \mu R^3 (1-x)^3} \rh 
 \lh \frac{p}{\mu} + \frac{\ve}{\mu} \rh, }
\ea
\label{4.2}
\ee
with 
\be
\frac{\ve}{\mu} = 1 - \frac{1}{4\pi \mu R^3 (1 - x)^2}\, \frac{d\cM}{dx}, \hs{2} 
\frac{p}{\mu} = - \lh \frac{\ve}{\mu} \rh^{-\ag}.
\label{4.3}
\ee
Large-$r$ solutions are now constructed by power series in $x$:
\be
\cM = \sum_{n \geq 0} \frac{m_n}{n!}\, x^n, \hs{2} \ve = \sum_{n \geq 0} \frac{\ve_n}{n!}\, x^n, \hs{2}
p = \sum_{n \geq 0} \frac{p_n}{n!}\, x^n.
\label{4.4}
\ee
Substitution into the equations (\ref{4.3}) allows one to compute the coefficients to arbitrary order. 
Results for the first 4 coefficients in each expansion are collected in appendix A. To obtain these
results equation (\ref{3.5}) is used to relate $m_0 = \cM(R)$ and $R$: 
\be
\frac{2 G m_0}{R} = 1 - \frac{8\pi G\mu}{3}\, R^2.
\label{4.7}
\ee
It is convenient to express all results in terms of two dimensionless variables characterizing 
the theory, the exponent $\ag$ and

\be
y = 8 \pi G \mu R^2.
\label{4.8}
\ee
For the coefficients of the energy density we then get:
\be
\ba{ll}
\dsp{ \frac{\ve_0}{\mu}\; =  }& \dsp{ \hs{-.2} y^{1/\ag}, \hs{2} 
\frac{\ve_1}{\mu} = \frac{1}{\ag}\, y^{1/\ag} \lh 3 - y^{1 + 1/\ag} \rh, }\\
 \\
\dsp{ \frac{\ve_2}{\mu}\; = }& \dsp{ \hs{-.2} \frac{y^{1/\ag}}{3\ag^2}  \lh  \frac{3 - y^{1 + 1/\ag}}{1 - y^{1 + 1/\ag}} \rh
 \left[ 9 - 7 \ag - (19 + 8\ag) y^{1 + 1/\ag} + (6 + 3\ag) y^{2+2/\ag} \right], }\\
 \\
\dsp{  \frac{\ve_3}{\mu}\; = }& \dsp{ \hs{-.2} \frac{y^{1/\ag}}{15 \ag^3 \lh 1 - y^{1 + 1/\ag} \rh} \left[ 
 - 405 + 945 \ag - 630 \ag^2 + \lh 2517 - 282 \ag  + 30 \ag^2 \rh y^{1 + 1/\ag} \rd }\\ 
 & \\
 & \dsp{ - \lh 4215 + 3035 \ag + 660 \ag^2 \rh y^{2 + 2/\ag} + \lh 2839 + 2937 \ag + 780 \ag^2 \rh 
 y^{3 + 3/\ag} }\\ 
 & \\
 & \dsp{ \ld - \lh 836 + 958 \ag + 270 \ag^2 \rh y^{4+4/\ag} + \lh 90 + 105 \ag + 30 \ag^2 \rh 
 y^{5 + 5/\ag} \right], }
\ea
\label{4.10}
\ee
Observe, that for the energy density to decrease with distance we must require $\ve_1 \geq 0$ or
\[
y^{1+ 1/\ag} \leq 3.
\]
The corresponding coefficients of the pressure are
\be
\ba{ll}
\dsp{  \frac{p_0}{\mu}\; = }& \dsp{ \hs{-.2} - \frac{1}{y}, \hs{2} 
\frac{p_1}{\mu} = \frac{1}{y} \lh 3 - y^{1 + 1/\ag} \rh, }\\
 & \\
\dsp{ \frac{p_2}{\mu}\; = }& \dsp{ \hs{-.2} \frac{1}{3\ag y} \lh  \frac{3 - y^{1 + 1/\ag}}{1 - y^{1 + 1/\ag}} \rh  
 \left[ - 16 \ag - (7 - 4 \ag) y^{1 + 1/\ag} + 3 y^{2+2/\ag} \right], }\\
 & \\
\dsp{  \frac{p_3}{\mu}\; = }& \dsp{ \hs{-.2} \frac{1}{15 \ag^2 y \lh 1 - y^{1 + 1/\ag} \rh} \left[ - 1980 \ag^2 
 + \lh 351 - 732 \ag + 1740 \ag^2 \rh y^{1 + 1/\ag} \rd }\\
 & \\
 & \dsp{ - \lh 885 - 415 \ag + 540 \ag^2 \rh y^{2 + 2/\ag} + \lh 739 + 117 \ag + 60 \ag^2 \rh y^{3 + 3/\ag} }\\
 & \\
 & \dsp{ \ld - \lh 251 + 103 \ag \rh y^{4 + 4/\ag} + \lh 30 + 15 \ag \rh y^{5 + 5/\ag} \right]. }
\ea
\label{4.10a}
\ee 

\bc
\scalebox{0.45}{\includegraphics{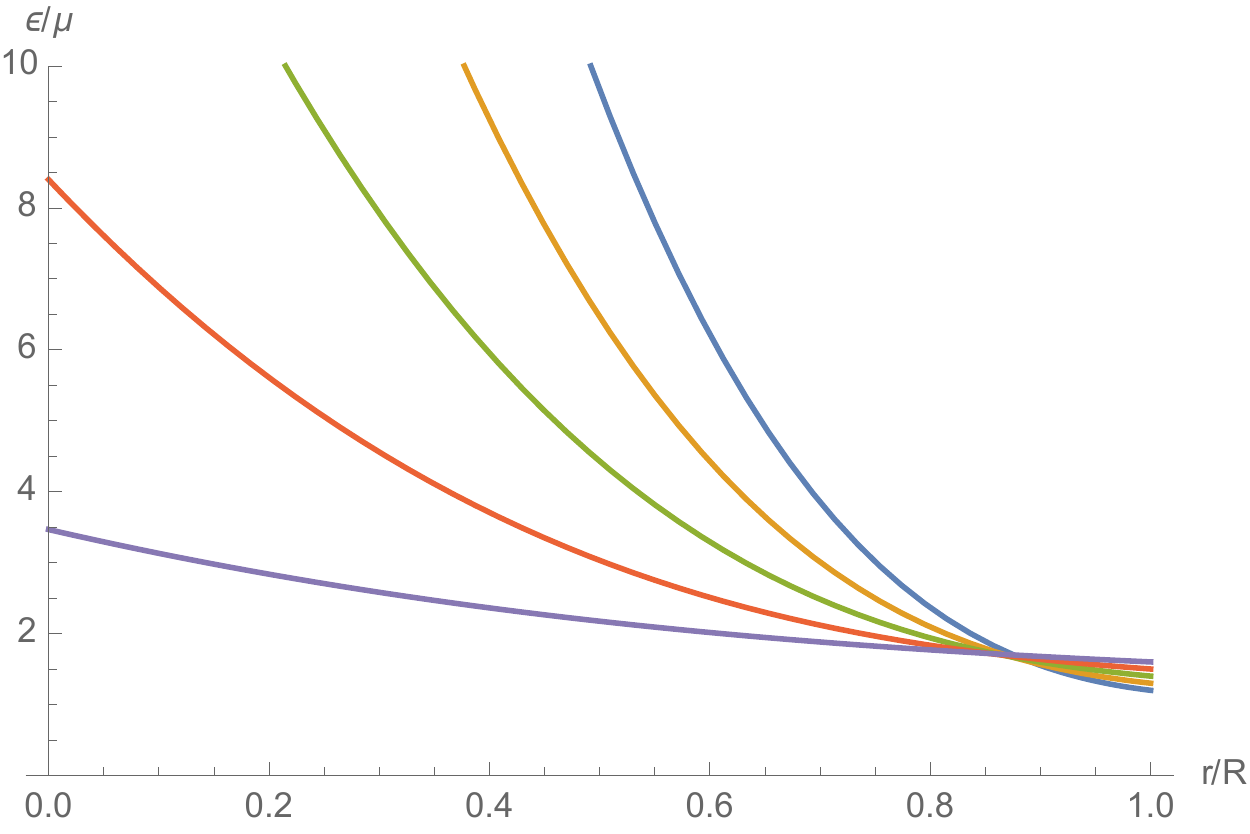}} 
\vs{2}

{\footnotesize Fig.\ 4.1: $\ve/\mu$ to order $x^3$ vs.\ $r/R$ for $\ag = 1 $ \\ 
      and from right to left:  $y=(1.2,1.3,1.4,1.5,1.6)$. }
\ec
\vs{1}

\bc
\scalebox{0.45}{\includegraphics{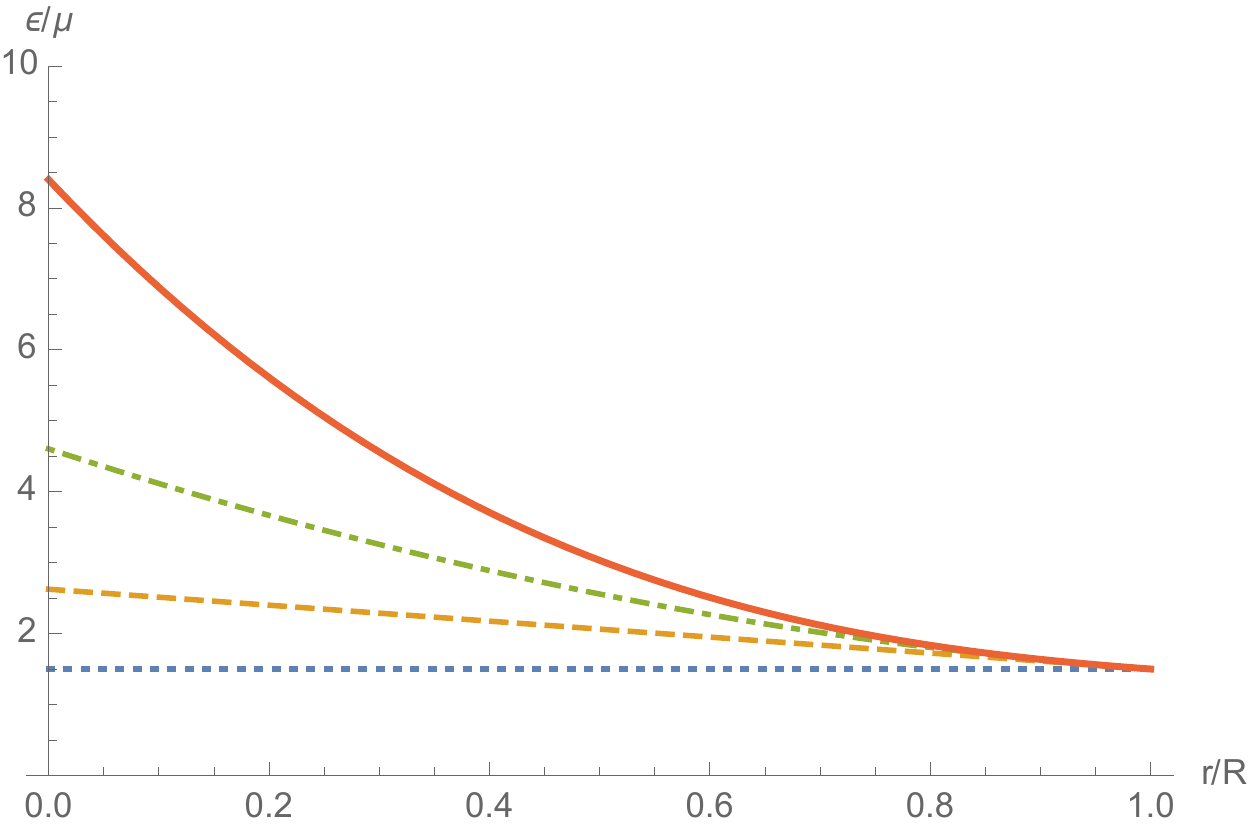}} 
\vs{2}

{\footnotesize Fig.\ 4.2: $\ve/\mu$ vs.\ $r/R$ for $\ag = 1$ and $y = 1.5$ to order $x^n$\\
   with from bottom to top $n = (0,1,2,3)$.}
\ec
\vs{1}

\nit
In figure 4.1 we show the results for the energy density $\ve/\mu$ as a function of $r/R$ 
to 3rd order in $x$ for the value $\ag = 1$ and various values of $y$. Similar figures for smaller 
values of $\ag$ are collected in appendix B. For the curve with $\ag = 1$ and $y= 1.5$ 
we also show separately the 0th-, 1st-, 2nd- and 3rd-order result for $\ve/\mu$ in figure 4.2. 
For this case the results are seen to converge well in the domain of large-$r$. 
\vs{2}

\nit
{\bf \large 5.\ Newtonian regime} 
\vs{1}

\nit
In the small-$r$ regime we can find a solution of the modified TOV equation using the newtonian 
approximation \ct{El-Zant2015}, in which it is assumed that $|p| \ll \mu \ll \ve$ and 
\be
\frac{2G \cM}{r} \ll 1, \hs{1} \mbox{whilst} \hs{1} \frac{\cM}{4\pi \mu r^3} \gg 1.
\label{5.1}
\ee
Thus close to the center of the halo $\cM$ is to grow faster than $r$ and slower than $r^3$
with increasing distance. In this approximation the modified TOV equation reduces to 
the condition for newtonian hydrostatic equilibrium:
\be
\frac{p'}{\ve} = - \frac{G\cM}{r^2}.
\label{5.2}
\ee
Using the gCg equation of state it follows that 
\be
G \cM = \frac{\ag r^2}{1+\ag}\, \frac{d}{dr} \left[ \lh \frac{\ve}{\mu} \rh^{-(1 + \ag)} \right].
\label{5.3}
\ee
Differentiating once more with respect to $r$ this results in a differential equation for 
the energy density:
\be
G \cM' \simeq 4\pi G r^2 \ve = \frac{\ag}{1+\ag}\, \frac{d}{dr} 
 \left\{ r^2 \frac{d}{dr} \left[ \lh \frac{\ve}{\mu} \rh^{-(1 + \ag)} \right] \right\},
\label{5.4}
\ee
with the solution 
\be
\frac{\ve}{\mu} = \lh \frac{r}{r_c} \rh^{- 2/(2+\ag)}, \hs{2} 2 \pi G \mu r_c^2 = \frac{\ag(4+3\ag)}{(2+\ag)^2}.
\label{5.5}
\ee
For the effective mass function this implies
\be
\frac{G\cM(r)}{r_c} = \frac{\ag (4 + 3\ag)}{(1+\ag)(2+\ag)} \lh \frac{r}{r_c} \rh^{(4 + 3\ag)/(2 + \ag)}, 
\label{5.6}
\ee
which satifies the initial assumptions (\ref{5.1}) for all positive values of $\ag$. Indeed, in the limit
$\ag \rightarrow 0$ it is seen to grow as $r^2$, whilst in the limit $\ag \rightarrow \infty$ it 
grows as $r^3$. Finally the assumption of small pressure: $|p| \ll \mu$, is satisfied in the domain 
$r \ll r_c$, with $|p| = 0$ in the center where the energy density $\ve$ diverges, although 
the mass function $\cM$ vanishes there and remains finite for all $r$ in the newtonian regime. 

In terms of the parameter $y$ introduced in (\ref{4.8}) the expression (\ref{5.5}) for the energy 
density can be recast in the form
\be
\frac{\ve}{\mu} = \lh \frac{r}{R} \rh^{-2/(2+\ag)} \left[ \frac{(2+\ag)^2 y}{4\ag (4+3\ag)} \right]^{-1/(2+\ag)}.
\label{5.7}
\ee
In figure 5.1 this expression is plotted and compared with the large-$r$ expansion for the cases 
$\ag = 1$ and three values of $y$.
\vs{1}

\bc
\scalebox{0.37}{\includegraphics{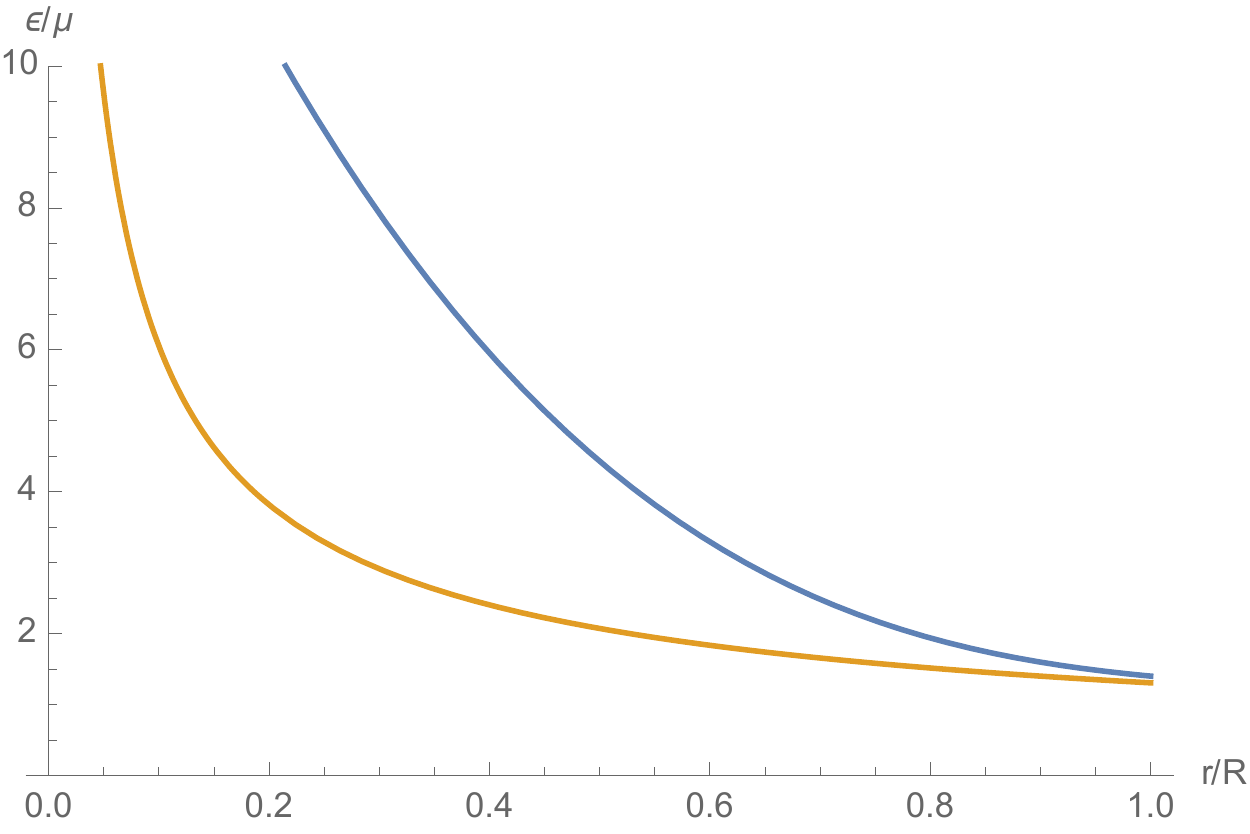}, \hs{1.5} 
 \includegraphics{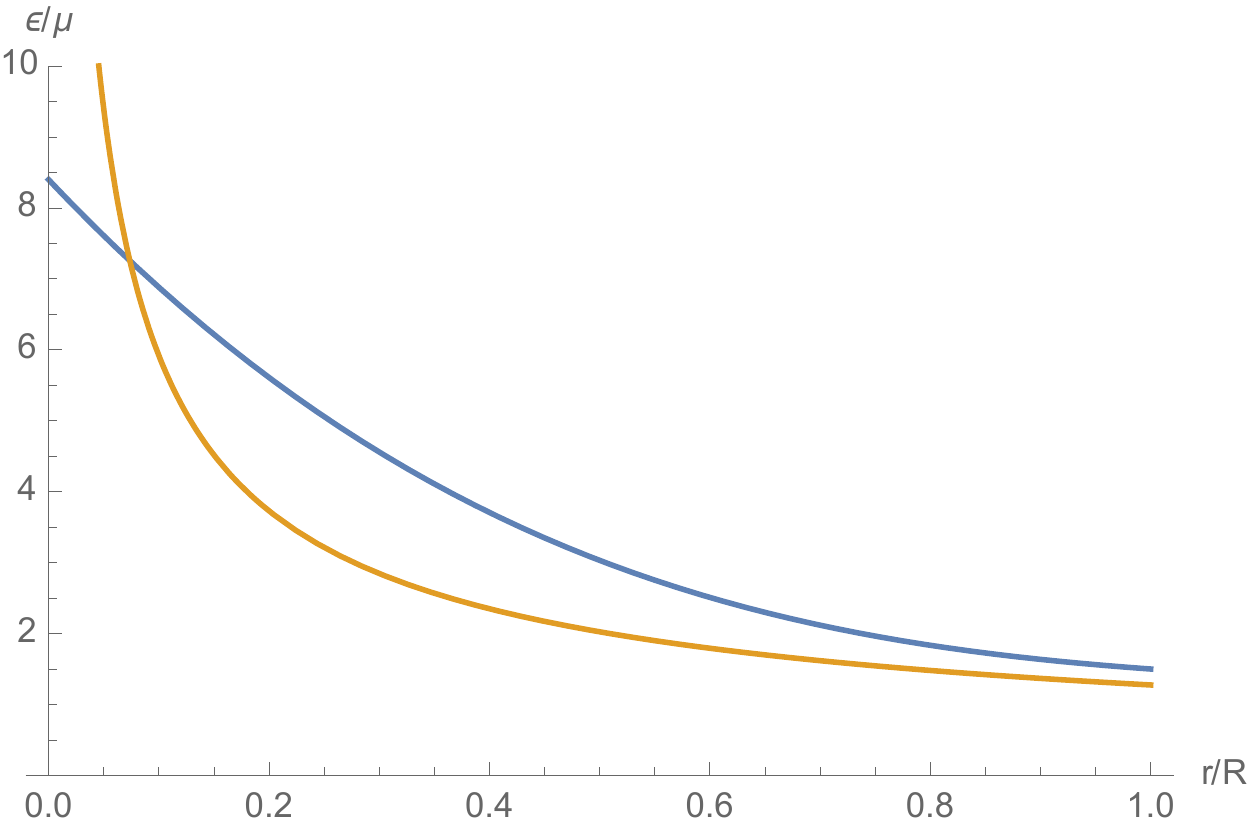}, \hs{1.5} \includegraphics{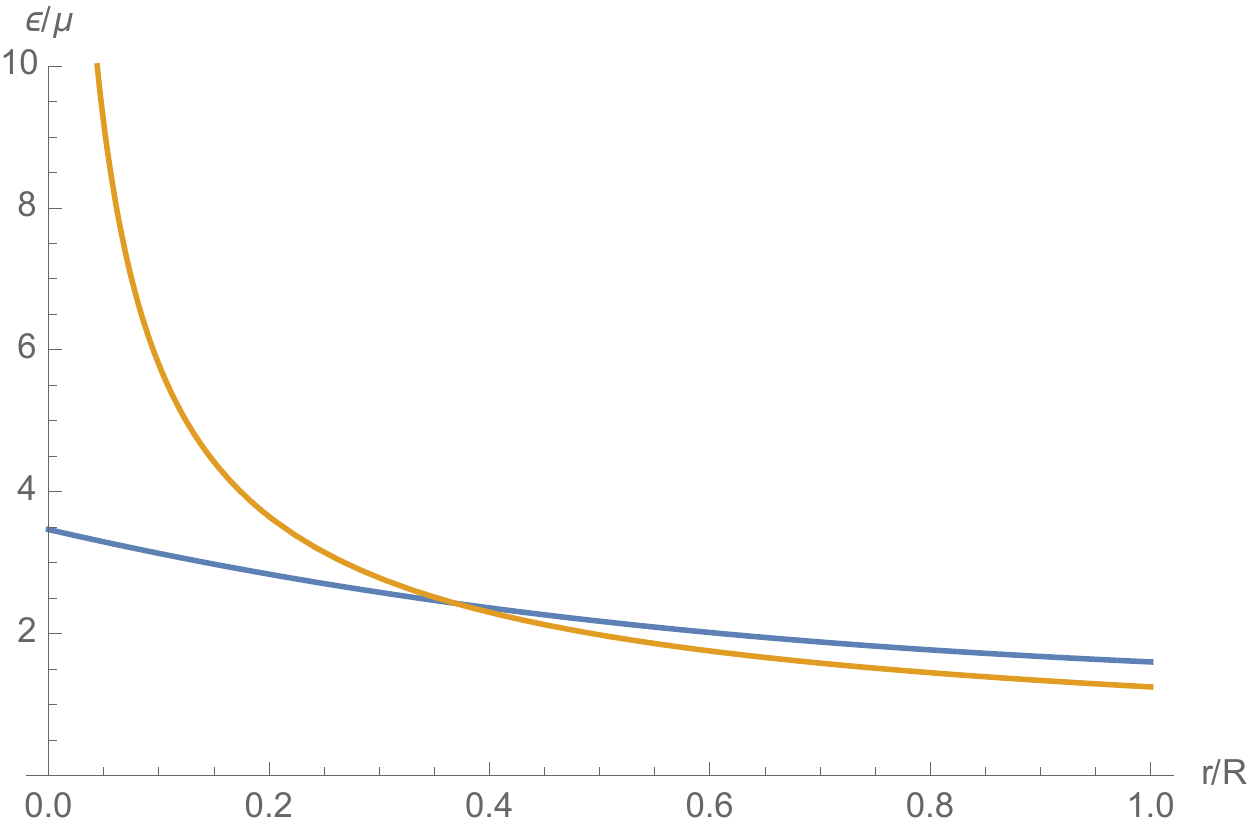}}
\vs{2}

{\footnotesize Fig.\ 5.1: Newtonian approximation (steeper orange curve) vs.\ large $r$-expansion 
 (flatter blue curve) \\ $\hs{2}$ of $\ve/\mu$ to order $x^3$ for $\ag = 1$ and $y = 1.4$ (left), $y = 1.5$ (middle) 
 and $y = 1.6$ (right). }
\ec
\vs{1}

\nit
These figures show that agreement between the two approximations in the large-$r$ region is quite good,
especially for the lowest value of $y$; this can be explained as 
\[
\lh \frac{r_c}{R} \rh^2 = \frac{4\ag (4 + 3\ag)}{(2+\ag)^2 y},
\]
and this ratio is close to 2 for the values of $\ag$ and $y$ used in the plots. For these and similar parameter 
values the whole region inside the horizon $r \leq R$ is inside the domain of the newtonian approximation, 
and it should get better the smaller the values of $y$. The figures also show that the large-$r$ expansion 
overestimates the small-$r$ values for the smaller value of $y$, while underestimating them for the larger 
$y$ value. The same tendencies hold for lower values of $\ag$, although the range of validity of the 
large-$r$ expansion is considerably more restricted there, as shown in appendix B.

\np
\nit
{\large \bf 6.\ Models with $\ag > 1$}
\vs{1}

\nit
So far we have considered models with $\ag \leq 1$ on the assumption that there is a universal 
bound on the speed of sound $c_s^2 \leq 1$. However, such a bound may be too strong in the
presence of a horizon. In fact for a radially decreasing energy density in the domain $r \leq R$ 
equation (\ref{2.5}) only requires 
\be
\ag \leq \frac{\ag}{c_s^2(R)} = \lh \frac{\ve_0}{\mu} \rh^{1 + \ag} = y^{1 + 1/\ag}.
\label{s6.1}
\ee
A natural boundary condition is to let $c_s$ take its maximal value on the horizon: 
$c_s(R) = 1$, which happens if the parameters $\ag$ and $y$ are related by
\be
\ag = y^{1 + 1/\ag}.
\label{s6.2}
\ee
As the asymptotic energy density $\ve_0/\mu = y^{1/\ag} > 1$ it follows that in 
this case also $\ag > 1$.
\vs{1}

\bc
\scalebox{0.5}{\includegraphics{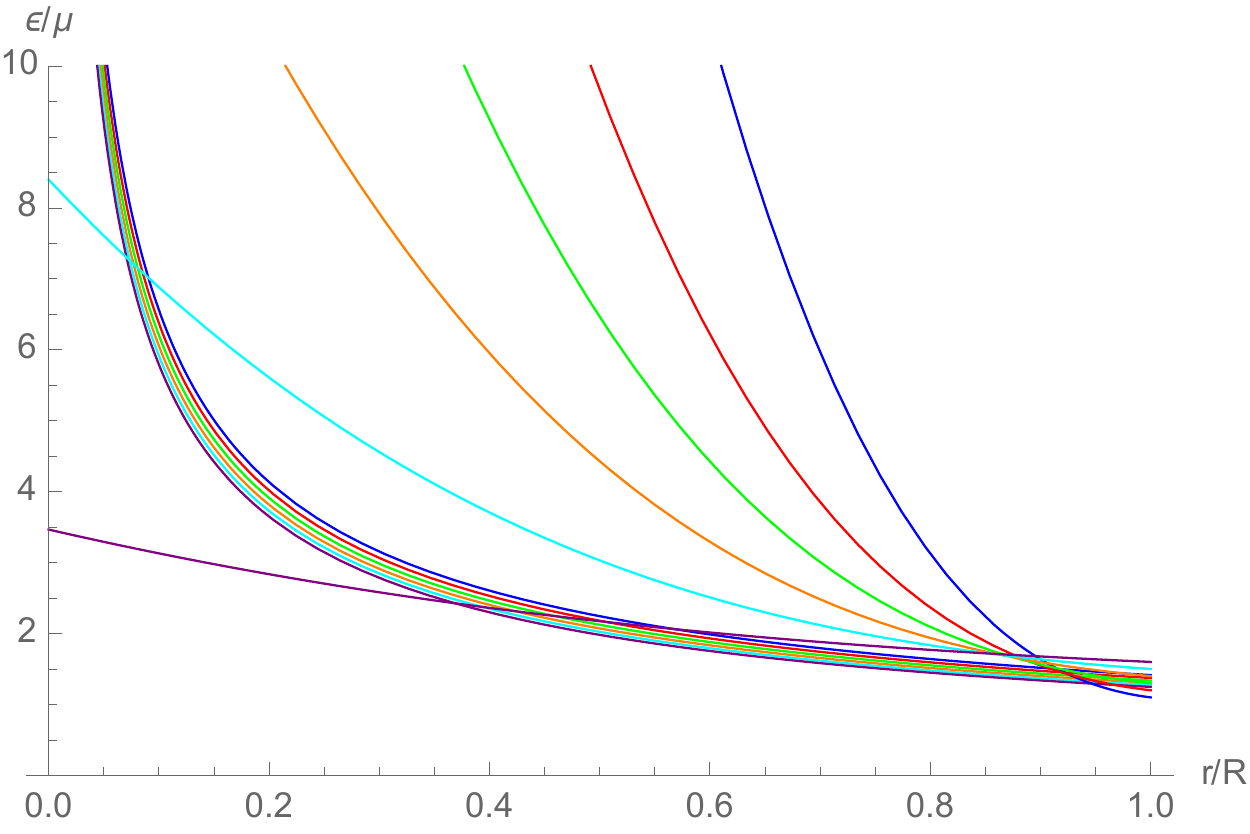}}
\vs{2}

{\footnotesize Fig.\ 6.1: Radial profile of the energy density $\ve/\mu$ for $c_s(R) = 1$ 
   in the 3rd-order large-$r$ and newtonian approximations; from right to left 
   $\ag = (1.2,1.4,1.6,1.8,2.0)$.}
\ec
\vs{1}

\nit
In fig.\ 6.1 we have plotted the energy density profiles for these models taking values of $\ag$ 
ranging from 1.2 to 2.0. The narrow bundle of lines represents the newtonian approximations, 
the wider bundle of lines with the lower asymptotic values of $\ve_0/\mu$ represents the 
large-$r$ approximations, which are relevant for $r$-values close to $R$. Even there the 
newtonian regime seems to be close to the true solution, which can be traced to the fact that 
in all these cases $R < r_c$, the more so for larger values of $\ag$. The exact asymptotic 
values of $\ve_0/\mu$ in the figure are all in the range 1.08 - 1.26,  close to unity.
\vs{2}

\nit
{\large \bf 7.\ Cosmological inference}
\vs{1}

\nit
Having worked out halo profiles predicted by gCg models, we can actually take input from 
cosmological data on dark matter and dark energy to fix some parameters. The parameter $\ve_0$ 
is the total asymptotic energy density near the de Sitter horizon. It is composed of both a dark-matter
and a dark-energy like component. We first have to determine how to separate these components 
for the gCg cosmology discussed here. Consider the spatial line element at fixed $t$:
\be
ds^2 = \frac{dr^2}{1 - \frac{2G\cM}{r} - \frac{8\pi G\mu}{3}\, r^2} + r^2 d\Og^2.
\label{d.1}
\ee
Asymptotically this behaves like the Schwarzschild-de Sitter line element at fixed $t$, with 
$\cM(R) = m_0$ the mass inside the sphere within the horizon, and $\mu$ the asymptotic 
cosmological constant. This suggests we interpret the asymptotic energy density $\ve_0$ as 
composed of a dark matter and a dark energy component such that 
\be
\frac{\ve_{de\, 0}}{\mu} = 1, \hs{2} \frac{\ve_{dm\, 0}}{\mu} = y^{1/\ag} - 1. 
\label{d.2}
\ee
Observations of the CMB \ct{PlanckCollaboration2015} give the ratio in the early universe to be 
\be
\frac{\ve_{dm}}{\ve_{de}} = 0.39.
\label{d.3}
\ee
Associating this value to our asymptotic expressions we get 
\be
y = (1.39)^{\ag}.
\label{d.4}
\ee
Thus $y$ varies between $y = 1$ for $\ag = 0$ to $y = 1.39$ for $\ag = 1$. These results are 
in the range previously considered. It appears that the results for values $\ag > 1$ do not fit 
the relation (\ref{s6.2}) very well. 
\vs{0.5} 

\bc
\scalebox{0.25}{\includegraphics{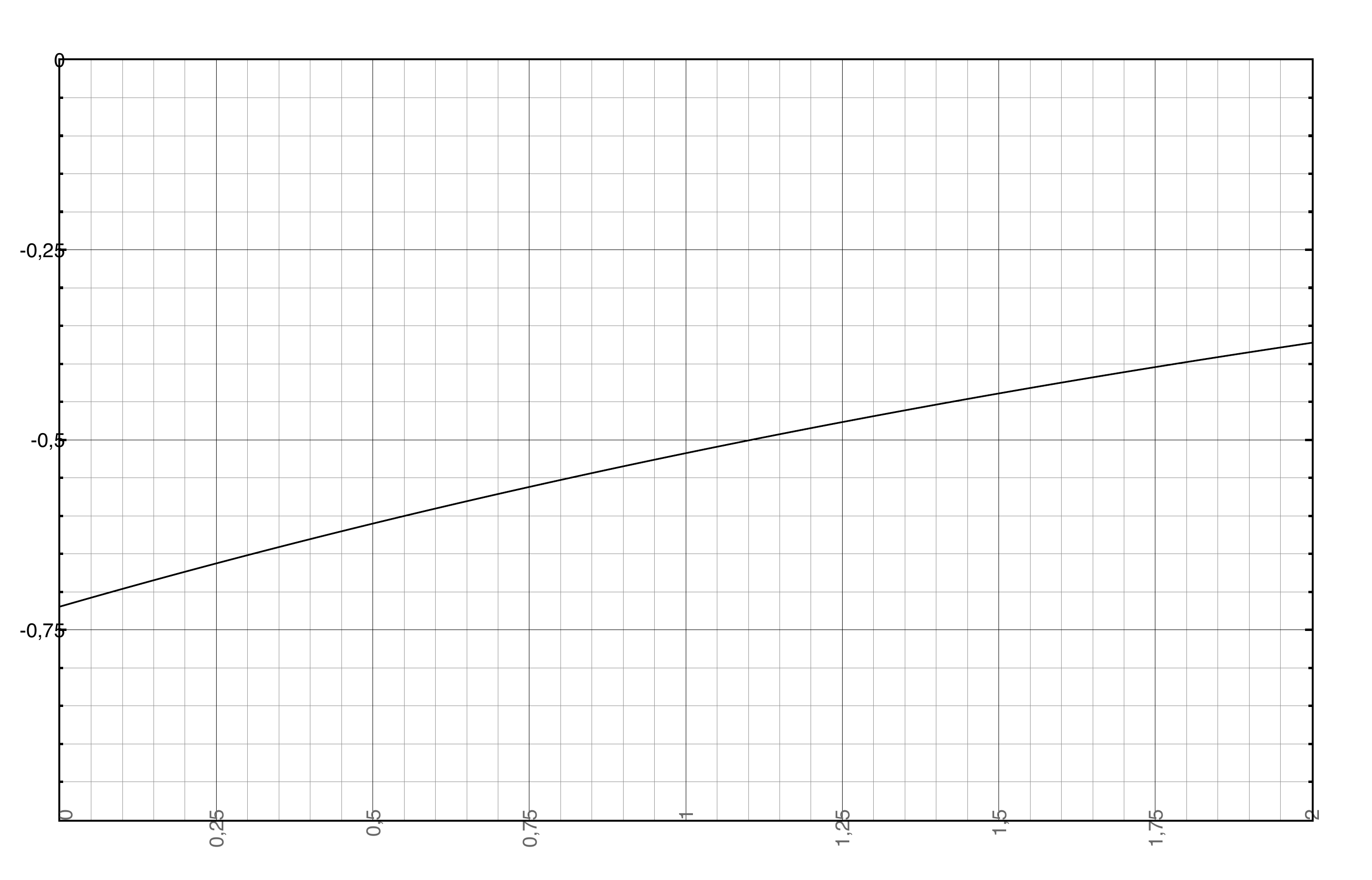}}
\vs{2}

{\footnotesize Fig.\ 7.1: $w_0$ vs.\ $\ag$ for the estimated asymptotic dark matter and energy 
densities \\ inferred from CMB data. }
\ec
\vs{0.5}

\nit
Similarly we can determine the asymptotic equation of state parameter for the gCg:
\be
w_0 = - \frac{1}{y^{1 + 1/\ag}},
\label{d.5}
\ee
with values between $w_0 = -0.72$ for $\ag = 0$ and $w_0 = - 0.52$ for $\ag = 1$, as shown 
in fig.\ 7.1. Note that some strong upper limits on $\ag$ were suggested in refs.\ 
\ct{Sandvik2002, Park2009,Aurich2017}, but these seem difficult to reconcile with the large-$r$ 
results (\ref{4.10}).
\vs{2}

\nit
{\large \bf 8.\ Space-time geometry and rotation profiles} 
\vs{1}

\nit
The space-time geometry in the spherically symmetric gCg halo is that of the line element (\ref{3.1}),
where $\cB(r)$ is given by (\ref{3.4}) and $\cA(r)$ is the solution of equation (\ref{3.6}):
\be
\cA(r) = \left[ 1 - \lh \frac{\ve}{\mu} \rh^{-(1 + \ag)} \right]^{\frac{-2\ag}{1 + \ag}},
\label{6.1}
\ee
where the constant of integration has been fixed so that $\cA(0) = 1$. With this choice the space-time 
is flat in the center of the halo. The line element (\ref{3.1}) implies for light-like radial geodesics 
\be
\lh \frac{dr}{dt} \rh^2 = \cA(r) \cB^{-1}(r) \hs{1} \stackrel{r = R}{\longrightarrow} \hs{1} 0,
\label{6.1.1}
\ee
showing explicitly the existence of the horizon at $R$ such that $\cB^{-1}(R) = 0$. In the newtonian 
regime the expressions take the approximate form
\be
\ba{l}
\dsp{ \cA_{newt} = \left[ 1 - \lh \frac{r}{r_c} \rh^{\frac{2+2\ag}{2+\ag}} \right]^{\frac{-2\ag}{1 + \ag}}, }\\
 \\
\dsp{ \cB_{newt} = \left[ 1 - \frac{2\ag(4+3\ag)}{(1+\ag)(2+\ag)}\, \lh \frac{r}{r_c} \rh^{\frac{2+2\ag}{2+\ag}}
 - \frac{4\ag(4+3\ag)}{3(2+\ag)^2}  \lh \frac{r}{r_c} \rh^2 \right]^{-1}. }
\ea
\label{6.2}
\ee
Note the limits
\be
\ba{l}
\dsp{ \ag = 0: \hs{1} \cA_{newt} = \cB_{newt} = 1, }\\
 \\
\dsp{ \ag = 1: \hs{1} \cA_{newt} = \left[ 1 - \lh \frac{r}{r_c} \rh^{4/3} \right]^{-1}, \hs{1}
\cB_{newt} = \left[ 1 - \frac{7}{3} \lh \frac{r}{r_c} \rh^{4/3} - \frac{28}{27} \lh \frac{r}{r_c} \rh^2 \right]^{-1}. }
\ea
\label{6.3}
\ee
Geodesic orbits for testmasses $m$ in a spherically symmetric space-time (\ref{3.1}) are planar, which 
we will take to be the equatiorial plane $\thg = \pi/2$.  In addition they are characterized by two constants 
of motion: the specific energy $\eta = E/m$ determined by the time dilation factor
\be
\eta = \cA\, \frac{dt}{d\tau}, 
\label{6.4}
\ee
and the specific angular momentum $\ell = L/m$ determined by the rotational velocity
\be
\ell = r^2\, \frac{d\vf}{d\tau}.
\label{6.5}
\ee
Finally time-like line elements satisfy the hamiltonian constraint 
\[
\cA \lh \frac{dt}{d\tau} \rh^2 = 1 + \cB \lh \frac{dr}{d\tau} \rh^2 + r^2 \lh \frac{d\vf}{d\tau} \rh^2.
\]
In view of the preceeding results, for circular orbits with constant $r = \mfr$ this implies the relation 
\be
\eta^2 = \cA(\mfr) \lh 1 + \frac{\ell^2}{\mfr^2} \rh,
\label{6.6}
\ee
whilst the stability of such orbits requires vanishing radial acceleration:
\be
\ld \frac{d^2 r}{d\tau^2} \right|_{r = \mfr} = 0 \hs{1} \Rightarrow \hs{1} 
\frac{\ell^2}{\mfr^2} = \left[ \frac{-2 \ag r\, \frac{\ve'}{\mu}}{\frac{\ve}{\mu} \lh \lh \frac{\ve}{\mu} \rh^{1 + \ag} - 1 \rh 
 + \ag r\, \frac{\ve'}{\mu}} \right]_{r = \mfr}.
\label{6.7}
\ee
In the newtonian regime $\mfr < r_c$, cf.\ (\ref{5.5}), this becomes
\be
v^2(\mfr) = \frac{\ell^2}{\mfr^2} = \frac{4\ag \lh \frac{\mfr}{r_c} \rh^{\frac{2+2\ag}{2 +\ag}} }{
 (2 + \ag) - (2 + 3\ag) \lh \frac{\mfr}{r_c} \rh^{\frac{2 + 2\ag}{2 + \ag}} },
\label{6.8}
\ee
where $v(\mfr)$ is the orbital velocity. For example for $\ag = 1$, which is in the regime $\mfr < r_c$ all 
the way up to the horizon $R$, we get 
\be
v^2(\mfr) = \frac{4}{3} \lh \frac{\mfr}{r_c} \rh^{4/3} \frac{1}{1 - \frac{5}{3} \lh \frac{\mfr}{r_c} \rh^{4/3}}
 \simeq \frac{4}{3} \lh \frac{\mfr}{r_c} \rh^{4/3} \lh 1 +  \frac{5}{3} \lh \frac{\mfr}{r_c} \rh^{4/3} + ... \rh.
\label{6.9}
\ee
To lowest-order approximation it follows that 
\be
v(\mfr) = \frac{2}{\sqrt{3}} \lh \frac{\mfr}{r_c} \rh^{2/3}.
\label{6.10}
\ee
In this discussion we have implicitly assumed that the space-time described by the line 
element (\ref{6.1}) is stationary. However the existence of the horizon at $r = R$ where 
$\cB^{-1}(R) = 0$ and the related homogeneous cosmological space-times (\ref{2.9}) 
indicate a time-dependent expanding geometry. We can make this explicit by performing 
a co-ordinate transformation defined implicitly by two functions $G(r)$ and $K(r)$ which are 
solutions of the equations
\be
G = \sqrt{\cA + H^2 r^2}, \hs{2} \frac{rK_r}{K} = 1 - \sqrt{\frac{\cA \cB}{\cA + H^2 r^2}},
\label{6.12}
\ee
where $K_r = dK/dr$ and $H$ is the asymptotic Hubble constant 
\be
H = \sqrt{ \frac{8\pi G\mu}{3}}.
\label{6.13}
\ee
Introducing new time and radial co-ordinates $\tau$ and $\varrho$:
\be
d\tau = dt - \frac{Hr (1 - r K_r/K)}{G^2 - H^2 r^2}\, dr, \hs{2} \varrho = e^{- H\tau} \frac{K(r)}{r},
\label{6.14}
\ee
the line-element (\ref{6.1}) takes the form 
\be
ds^2 = - \gam^2(\varrho,\tau) d\tau^2 + e^{2H\tau} \kg^2(\varrho, \tau)) 
 \lh d\varrho^2 + \varrho^2 d\Og^2 \rh,
\label{6.15}
\ee
where $\gam$ and $\kg$ are given in terms of the solutions of eqs.\ (\ref{6.12}) as
\be
\gam(\varrho, \tau) = G(r), \hs{2} \kg(\varrho, \tau) = K(r).
\label{6.16}
\ee
This shows that $\gam$ and $\kg$ are in fact functions of a single variable $e^{H\tau} \varrho = K(r)/r$. 
Equation (\ref{6.15}) is to replace relation (\ref{2.9}) when taking a simple form of halo structure of dark 
matter into account. It also follows that orbits $r = \mfr$ are near-closed circular only in the newtonian limit 
where the period $T = 2\pi \mfr/v(\mfr)$ of the orbit is much smaller than the asymptotic Hubble time $1/H$. 
\vs{3}

\nit
{\large \bf 9.\ Summary  and discussion}
\vs{1}

\nit
In this paper we have analysed the structure and cosmological implications of spherical halos of a 
generalized Chaplygin gas. We have shown that a non-trivial spherically symmetric distribution of 
gCg creates a horizon at finite radial co-ordinate (but infinite proper distance) in agreement with 
\ct{Gorini2009}. The density of the gas decreases monotonically towards the horizon, but remains 
finite non-zero up to the largest distances. We have also found that in many cases the newtonian 
approximation for the structure of the halo works well, even though the space-time itself is 
characterized by a non-flat metric with coefficients (\ref{6.2}). It gives rise to a rather weak 
dependence on the radius of orbital velocities of test masses in circular orbits, equation (\ref{6.8}), 
for orbits much smaller than the horizon distance. 

For a realistic description of cosmological structures the model has to be extended in several 
ways. First, the observable universe contains a large number of clusters of galaxies with 
overlapping dark-matter halos within a single de Sitter-like horizon. Still, these would be 
expected to give rise to a universal asymptotic behaviour as described by our large-$r$ 
expansion in section 4. Second, in addition to dark matter and dark energy our universe also 
contains baryonic matter and radiative components; it may also contain additional dark-matter 
components. All of these have to be taken into account to get a realistic cosmology as 
discussed e.g.\ in \ct{Gorini2003, Gorini2008}.

Nevertheless taking into account such simplifications made here the generalized Chaplygin gas 
appears to offer a more flexible effective theory of dark energy and dark matter allowing for richer 
structures with varying dark matter as well as dark energy density than a simple cosmological 
constant, as in the $\Lb$CDM models. As such it can be of value in parametrizing the observed 
cosmological features of our universe.

\vs{3}
\nit
{\bf Acknowledgement} \\
For JwvH this work is part of the research programme of the Netherlands Research Organisation NWO-I. 
His work is supported by the Lorentz Foundation (Leiden). 

\np
\nit
{\large \bf Appendix A} 
\vs{1} 

\nit
The expansion coefficients in equations (\ref{4.4}) are related by the definition of $\cM$,
equation (\ref{3.3}), and the gCg equation of state (\ref{2.2}). As a result
\be
\ba{ll}
\dsp{ \frac{\ve_0}{\mu} = 1 - \frac{m_1}{4\pi \mu R^3}, }& 
\dsp{ \frac{\ve_1}{\mu} = - \frac{m_2 + 2m_1}{4\pi \mu R^3}, }\\
 & \\
\dsp{ \frac{\ve_2}{\mu} = - \frac{m_3 + 4m_2 + 6 m_1}{4\pi \mu R^3}, }& 
\dsp{ \frac{\ve_3}{\mu} = - \frac{m_4 + 6m_3 + 18 m_2 + 24 m_1}{4\pi \mu R^3}, }
\ea
\label{4.5}
\ee
and 
\be
\ba{l}
\dsp{ \frac{p_0}{\mu} = - \lh \frac{\ve_0}{\mu} \rh^{-\ag},  \hs{2}
 \frac{p_1}{\mu} = \ag \lh \frac{\ve_0}{\mu} \rh^{-(1+ \ag)} \frac{\ve_1}{\mu}, }\\
 \\
\dsp{ \frac{p_2}{\mu} = \ag  \lh \frac{\ve_0}{\mu} \rh^{-(1 + \ag)} \frac{\ve_2}{\mu} -
 \ag\, (1 + \ag)\, \lh \frac{\ve_0}{\mu} \rh^{-(2+\ag)} \lh \frac{\ve_1}{\mu} \rh^2, }\\
 \\
\dsp{ \frac{p_3}{\mu} = \ag \lh \frac{\ve_0}{\mu} \rh^{-(1+\ag)} \frac{\ve_3}{\mu} - 
 3 \lh 1 + \ag \rh \lh \frac{\ve_0}{\mu} \rh^{-(2+\ag)} \frac{\ve_1 \ve_2}{\mu^2} }\\
 \\
\dsp{ \hs{2.5} +\, \ag \lh 1 + \ag \rh \lh 2 + \ag \rh \lh \frac{\ve_0}{\mu} \rh^{-(3 + \ag)} 
 \lh \frac{\ve_1}{\mu} \rh^3. }
\ea
\label{4.6}
\ee
Using these results in the modified TOV equation (\ref{4.2}) we get in terms of $y = 8 \pi G \mu R^2$
\be
\ba{l} 
\dsp{ \frac{2G m_0}{R} = 1 - \frac{y}{3}, \hs{2} \frac{2Gm_1}{R} = y - y^{1 + 1/\ag}, }\\
 \\
\dsp{ \frac{2Gm_2}{R} = - 2 y + \lh 2 - \frac{3}{\ag} \rh y^{1 + 1/\ag} + \frac{1}{\ag}\, y^{2 + 2/\ag}, }\\
 \\
\dsp{ \frac{2Gm_3}{R} =\, 2y \lh 1 - y^{1+1/\ag} \rh }\\
 \\
\dsp{ \hs{4} -\, \frac{1}{3\ag^2} \lh \frac{3 - y^{1+1/\ag}}{1 - y^{1+1/\ag}} \rh \left[ (9-19 \ag)\, y^{1 + 1/\ag}
 - (19+20 \ag)\, y^{2 + 2/\ag} + (6 + 3\ag)\, y^{3 + 3/\ag} \right] }
\ea
\label{4.9}
\ee

\np
\nit
{\large \bf Appendix B}
\vs{1}

\nit
Here we show the 3rd-order large-$r$ expansion of the energy density $\ve/\mu$ as a function of $r/R$ for 
values of the Chaplygin exponent $\ag = (0.8, 0.6, 0.4)$. For the smallest values of $\ag$ the expansion 
seems to be reliable only at the very high end of $r$-values.

\bc
\scalebox{0.45}{\includegraphics{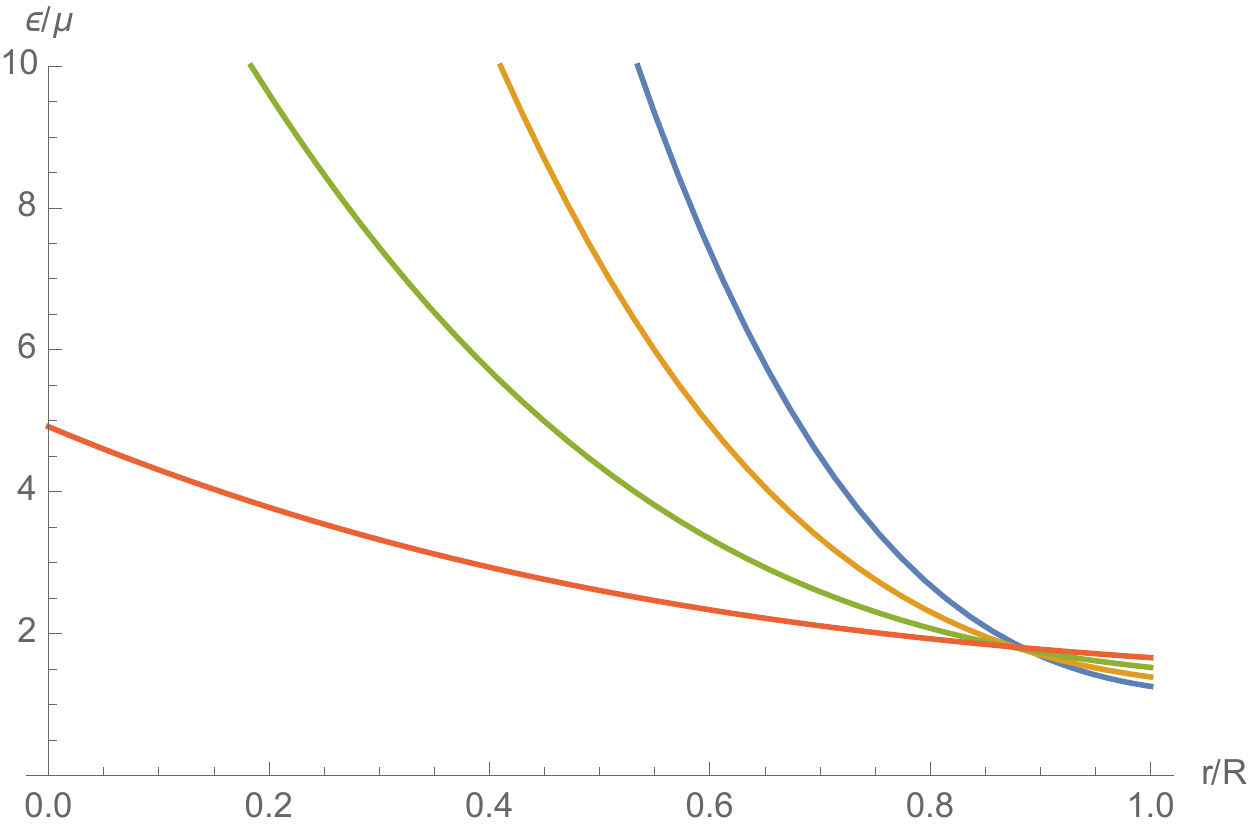}}
\vs{2}

{\footnotesize Fig.\ B.1: $\ve/\mu$ vs.\ $r/R$ for $\ag = 0.8$ and from right to left $y = (1.2,1.3,1.4,1.5)$. }
\vs{2}

\scalebox{0.45}{\includegraphics{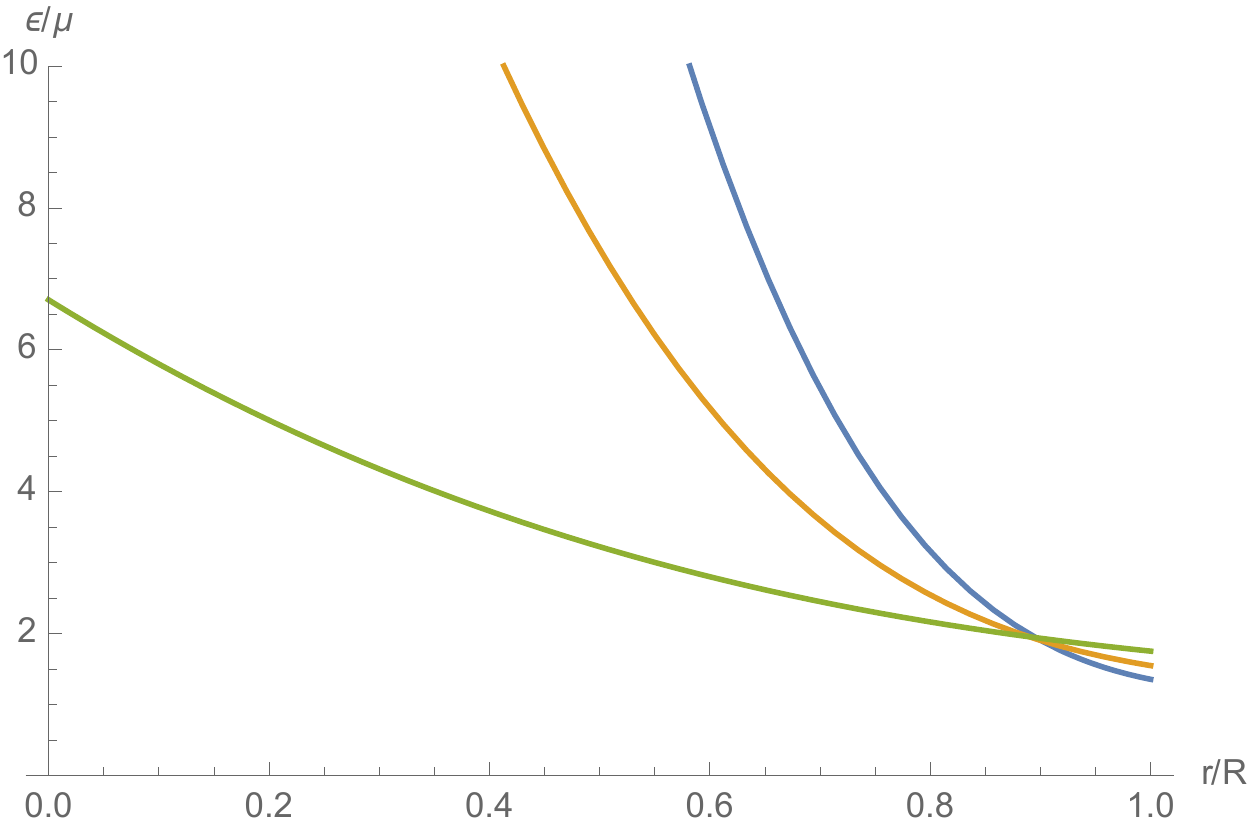}}
\vs{2}

{\footnotesize Fig.\ B.2: $\ve/\mu$ vs.\ $r/R$ for $\ag = 0.6$ and from right to left $y = (1.2, 1.3,1.4)$. }
\vs{2} 
 
\scalebox{0.45}{\includegraphics{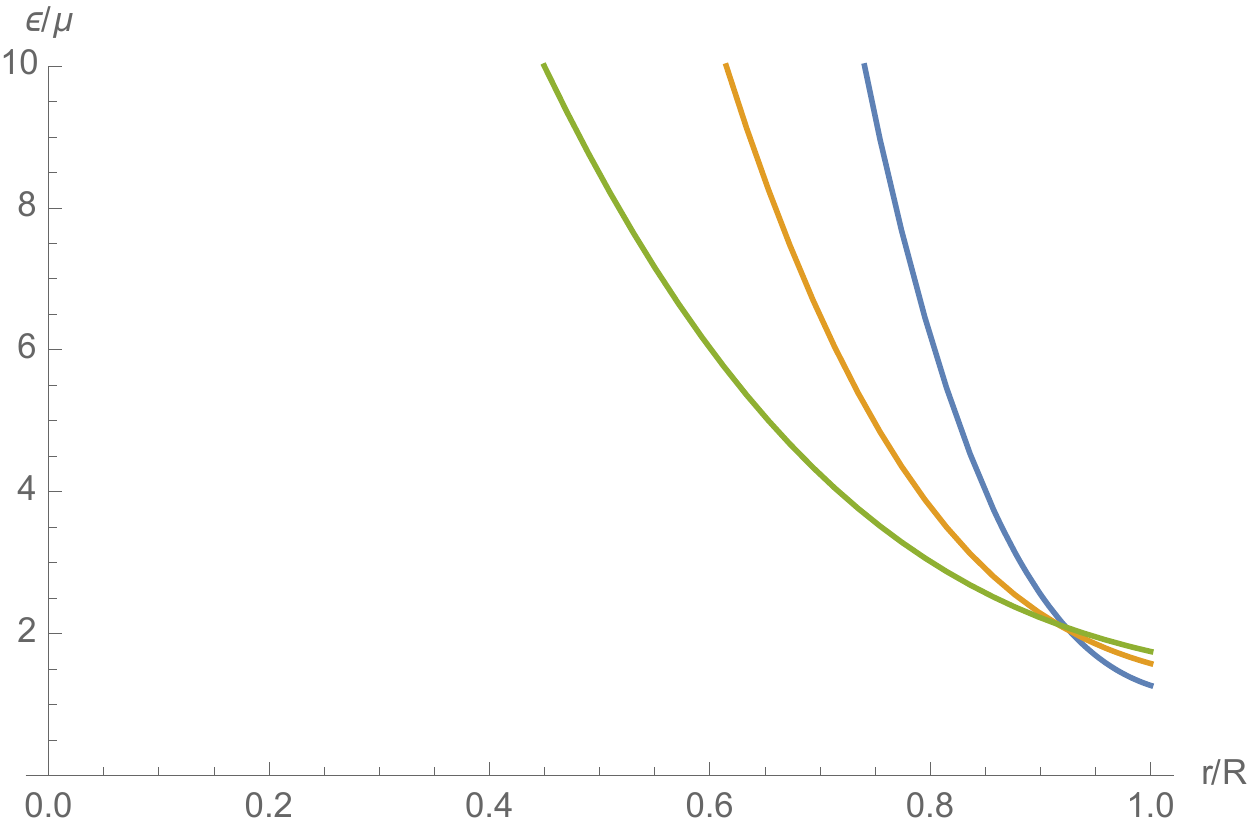}}
\vs{2}

{\footnotesize Fig.\ B.3: $\ve/\mu$ vs.\ $r/R$ for $\ag = 0.4$ and from right to left $y = (1.1,1.2, 1.25)$. }
\vs{2}
\ec

\nit
We also show the comparison of the large-$r$ expansion with the newtonian approximation for the energy 
density for the very small value $\ag = 0.25$ and two values of $y$ for which $r_c < R$.
\vs{1}

\bc 
\scalebox{0.4}{\includegraphics{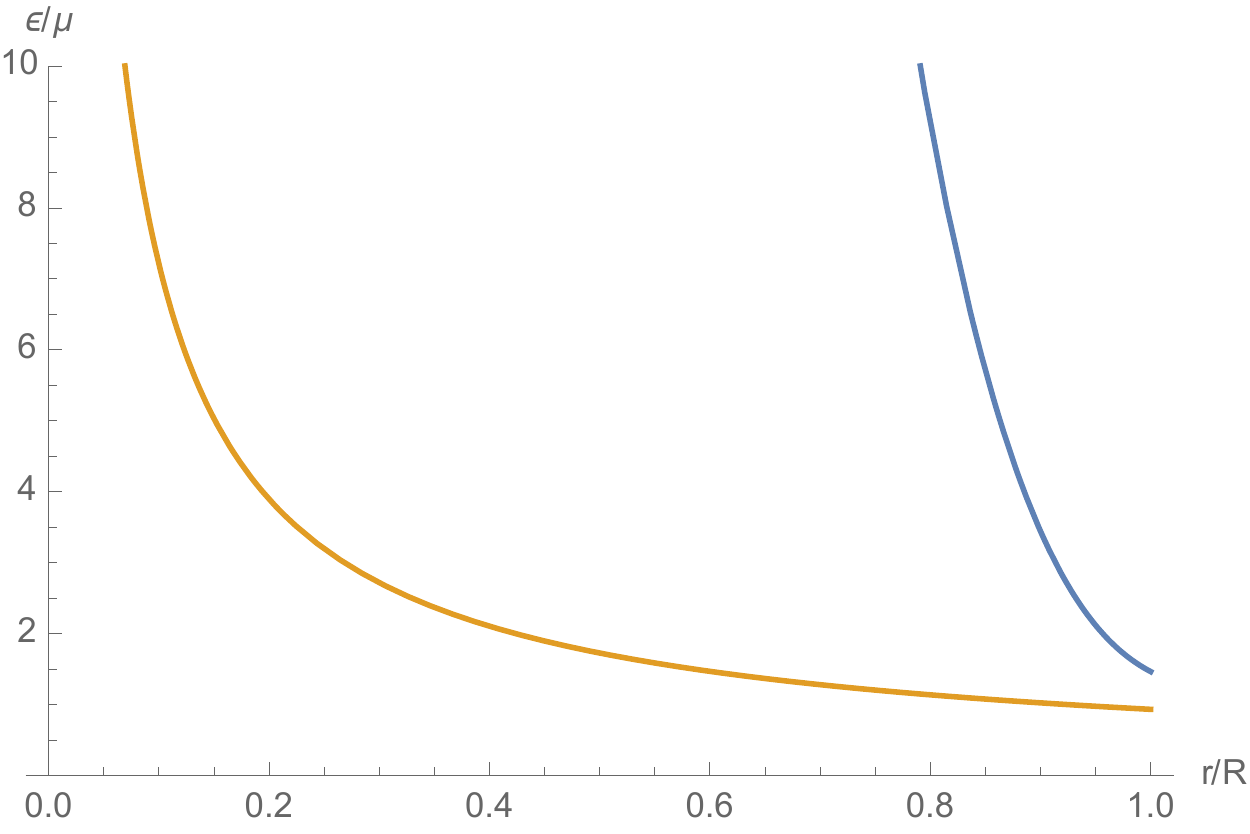}, \hs{1.5}
 \includegraphics{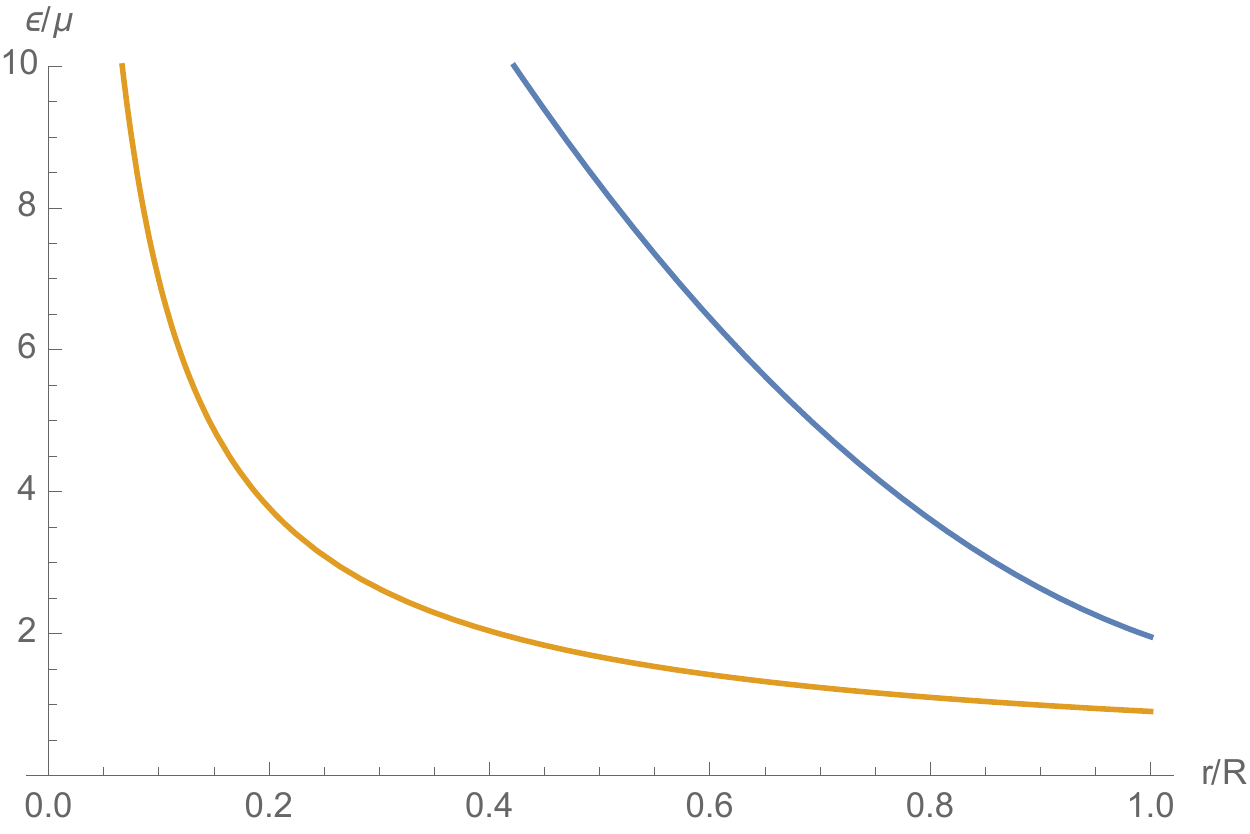}}
\vs{2}

{\footnotesize Fig.\ B.4 Comparison of newtonian approximation (lower curve) 
 vs.\ large-$r$ expansion (upper curve) \\
 for $\ag = 0.25$ and $y = 1.1$ (left), $y = 1.1825$ (right). }
\ec
\vs{1}

\nit
In both cases the newtonian approximation underestimates the large-$r$ value of the energy density, the
more so for larger $y$, whereas the large-$r$ expansion diverges much too fast for values of $r$ away 
from the horizon $r = R$.

\bibliographystyle{newutphys}
\bibliography{cgc_bib}

\providecommand{\href}[2]{#2}\begingroup\raggedright\begin{thebibliography}{10}

\bibitem{Zwicky1933}
F.~Zwicky, ``Die rotverschiebung von extragalaktischen nebeln,'' {\em Helvetica
  Physica Acta} {\bf 6} (1933)  110.

\bibitem{Bosma1979}
{A.\ Bosma} and {P.C.\ van der Kruit}, ``The local mass-to-light ratio in
  spiral galaxies,'' {\em Astron. Astrophys.} {\bf 79} (1979)  281--286.

\bibitem{Rubin1980}
V.~C. Rubin, N.~Thonnard, and J.~{Ford, W. K.}, ``{Rotational properties of 21
  SC galaxies with a large range of luminosities and radii, from NGC 4605 /R =
  4kpc/ to UGC 2885 /R = 122 kpc/},''
  \href{http://dx.doi.org/10.1086/158003}{{\em Astrophys. J.} {\bf 238} (1980)
  471}.

\bibitem{Clowe2006}
{D.\ Clowe et al.}, ``{A direct empirical proof of the existence of dark
  matter},'' \href{http://dx.doi.org/10.1086/508162}{{\em Astrophys.J.} {\bf
  648} (2006)  L109--L113}, \href{http://arxiv.org/abs/astro-ph/0608407}{{\tt
  arXiv:astro-ph/0608407}}.

\bibitem{Riess1998}
{A.G.\ Riess et al.}, ``{Observational Evidence from Supernovae for an
  Accelerating Universe and a Cosmological Constant},''
  \href{http://arxiv.org/abs/astro-ph/9805201}{{\tt arXiv:astro-ph/9805201}}.

\bibitem{Perlmutter1999}
S.~Perlmutter, ``{Measurements of Omega and Lambda From 42 High-Redshift
  Supernovae},'' \href{http://dx.doi.org/10.1086/307221}{{\em Astrophys. J.}
  {\bf 517} (1999)  565--586},
  \href{http://arxiv.org/abs/astro-ph/9812133}{{\tt arXiv:astro-ph/9812133}}.

\bibitem{Astier2012}
P.~Astier and R.~Pain, ``{Observational evidence of the accelerated expansion
  of the universe},'' \href{http://dx.doi.org/10.1016/j.crhy.2012.04.009}{{\em
  Comptes Rendus Phys.} {\bf 13} (2012) no.~6-7, 521--538},
  \href{http://arxiv.org/abs/1204.5493}{{\tt arXiv:1204.5493 [astro-ph]}}.

\bibitem{PlanckCollaboration2015}
{P.\ Ade et al.}, ``Planck 2015 results,''
  \href{http://dx.doi.org/10.1051/0004-6361/201525830}{{\em Astron. Astrophys.}
  {\bf 594} (2016)  A13}, \href{http://arxiv.org/abs/1502.01589}{{\tt
  arXiv:1502.01589 [astro-ph]}}.

\bibitem{Steigman1985}
{G.\ Steigman} and {M.S.\ Turner}, ``Cosmological constraints on the properties
  of weakly interacting massive particles,'' {\em Nucl.\ Phys.} {\bf B253}
  (1985)  375.

\bibitem{Peter2012}
{A.H.G.\ Peter}, ``{Dark Matter: A Brief Review},''
  \href{http://dx.doi.org/10.1088/1742-6596/378/1/012012}{{\em \em Proc.\ Frank
  N. Bash Symposium} (2011)  }, \href{http://arxiv.org/abs/1201.3942}{{\tt
  arXiv:1201.3942 [astro-ph]}}.

\bibitem{Ringwald2016}
{A.\ Ringwald}, ``{Alternative dark matter candidates: Axions},''
  \href{http://arxiv.org/abs/1612.08933}{{\tt arXiv:1612.08933 [hep-ph]}}.

\bibitem{Wetterich1988}
C.~Wetterich, ``{Cosmology and the fate of dilatation symmetry},''
  \href{http://dx.doi.org/10.1016/0550-3213(88)90193-9}{{\em Nucl. Phys. B}
  {\bf 302} (1988)  668--696}.

\bibitem{Wetterich2002}
{C.\ Wetterich}, ``Quintessence - the dark energy in the universe?,'' {\em
  Space Sci.Rev.} {\bf 100} (2002)  195--206,
  \href{http://arxiv.org/abs/astro-ph/0110211}{{\tt arXiv:astro-ph/0110211}}.

\bibitem{Sahni2002}
{V.\ Sahni} and {Y.\ Shtanov}, ``{Braneworld models of dark energy},''
  \href{http://dx.doi.org/10.1088/1475-7516/2003/11/014}{{\em JCAP} {\bf 0311}
  (2003)  014}, \href{http://arxiv.org/abs/astro-ph/0202346v3}{{\tt
  arXiv:astro-ph/0202346v3}}.

\bibitem{Verlinde2016}
E.~P. Verlinde, ``{Emergent Gravity and the Dark Universe},'' {\em SciPost
  Phys.} {\bf 2} (2017)  16, \href{http://arxiv.org/abs/1611.02269}{{\tt
  arXiv:1611.02269 [hep-th]}}.

\bibitem{Haridasu2017}
{B.S.\ Haridasu}, {V.V.\ Lukovi{\'{c}}}, {R.\ D'Agostino}, and {N.\ Vittorio},
  ``{Strong evidence for an accelerating Universe},''
  \href{http://dx.doi.org/10.1051/0004-6361/201730469}{{\em Astron. Astrophys.}
  {\bf 600} (2017) no.~2015, L1}, \href{http://arxiv.org/abs/1702.08244}{{\tt
  arXiv:1702.08244 [astro-ph]}}.

\bibitem{Kamenshchik2001}
A.~Kamenshchik, U.~Moschella, and V.~Pasquier, ``{An alternative to
  quintessence},'' \href{http://dx.doi.org/10.1016/S0370-2693(01)00571-8}{{\em
  Phys. Lett. Sect. B} {\bf 511} (2001)  265--268},
  \href{http://arxiv.org/abs/gr-qc/0103004}{{\tt arXiv:gr-qc/0103004}}.

\bibitem{Bento2002}
M.~C. Bento, O.~Bertolami, and A.~A. Sen, ``{Generalized Chaplygin gas,
  accelerated expansion, and dark-energy-matter unification},''
  \href{http://dx.doi.org/10.1103/PhysRevD.66.043507}{{\em Phys. Rev. D} {\bf
  66} (2002)  43507}, \href{http://arxiv.org/abs/gr-qc/0202064}{{\tt
  arXiv:gr-qc/0202064}}.

\bibitem{Bilic2002}
N.~Bilic, ``{Unification of dark matter and dark energy: the inhomogeneous
  Chaplygin gas},'' \href{http://dx.doi.org/10.1016/S0370-2693(02)01716-1}{{\em
  Phys. Lett. B} {\bf 535} (2002)  17--21},
  \href{http://arxiv.org/abs/astro-ph/0111325}{{\tt arXiv:astro-ph/0111325}}.

\bibitem{Sandvik2002}
{H.B.\ Sandvik, M.\ Tegmark, M.\ Zaldarriaga and I.\ Waga}, ``{The End of
  Unified Dark matter?},'' {\em Phys.\ Rev.\ D} {\bf 69} (2004)  123524,
  \href{http://arxiv.org/abs/astro-ph/0212114v2}{{\tt
  arXiv:astro-ph/0212114v2}}.

\bibitem{alcaniz2003a}
J.~Abha~Dev, Alcaniz and D.~Jain, ``Cosmological consequences of a chaplygin
  gas dark energy,'' {\em Phys.\ Rev.\ D} {\bf 67} (2003)  023515,
  \href{http://arxiv.org/abs/astro-ph/0209379}{{\tt arXiv:astro-ph/0209379}}.

\bibitem{alcaniz2003b}
D.~J. Alcaniz, J.S. and A.~Dev, ``High-redshift objects and the generalized
  chaplygin gas,'' {\em Phys.\ Rev.\ D} {\bf 67} (2003)  043514,
  \href{http://arxiv.org/abs/astro-ph/0210476}{{\tt arXiv:astro-ph/0210476}}.

\bibitem{Gorini2003}
V.~Gorini, A.~Kamenshchik, and U.~Moschella, ``{Can the Chaplygin gas be a
  plausible model for dark energy?},''
  \href{http://dx.doi.org/10.1103/PhysRevD.67.063509}{{\em Phys. Rev. D} {\bf
  67} (2003) no.~6, 1--6}, \href{http://arxiv.org/abs/astro-ph/0209395}{{\tt
  arXiv:astro-ph/0209395}}.

\bibitem{Makler2003}
M.~Makler, S.~Q. {De Oliveira}, and I.~Waga, ``{Constraints on the generalized
  Chaplygin gas from supernovae observations},''
  \href{http://dx.doi.org/10.1016/S0370-2693(03)00038-8}{{\em Phys. Lett. Sect.
  B Nucl. Elem. Part. High-Energy Phys.} {\bf 555} (2003)  1--6},
  \href{http://arxiv.org/abs/astro-ph/0209486}{{\tt arXiv:astro-ph/0209486}}.

\bibitem{Gorini2008}
V.~Gorini, A.~Y. Kamenshchik, U.~Moschella, O.~F. Piattella, and A.~A.
  Starobinsky, ``{Gauge-invariant analysis of perturbations in Chaplygin gas
  unified models of dark matter and dark energy},''
  \href{http://dx.doi.org/10.1088/1475-7516/2008/02/016}{{\em J. Cosmol.
  Astropart. Phys.} {\bf 2008} (2008)  016},
  \href{http://arxiv.org/abs/0711.4242}{{\tt arXiv:0711.4242 [astro-ph]}}.

\bibitem{Gorini2009}
V.~Gorini, A.~Y. Kamenshchik, U.~Moschella, O.~F. Piattella, and A.~A.
  Starobinsky, ``{More about the Tolman-Oppenheimer-Volkoff equations for the
  generalized Chaplygin gas},''
  \href{http://dx.doi.org/10.1103/PhysRevD.80.104038}{{\em Phys. Rev. D - Part.
  Fields, Gravit. Cosmol.} {\bf 80} (2009)  1--8},
  \href{http://arxiv.org/abs/0909.0866}{{\tt arXiv:0909.0866 [astro-ph]}}.

\bibitem{Park2009}
{C-G.\ Park}, {J-C.\ Hwang}, {J.\ Park}, and {H.\ Noh}, ``Observational
  constraints on a unified dark matter and dark energy model based on
  generalized chaplygin gas,'' {\em Phys.\ Rev.\ D} {\bf 81} (2010)  063532,
  \href{http://arxiv.org/abs/0910.4202}{{\tt arXiv:0910.4202 [astro-ph]}}.

\bibitem{El-Zant2015}
A.~A. El-Zant, ``{Unified dark matter: Constraints from galaxies and
  clusters},'' \href{http://dx.doi.org/10.1093/mnras/stv1700}{{\em Mon. Not. R.
  Astron. Soc.} {\bf 453} (2015)  2250--2258},
  \href{http://arxiv.org/abs/1507.07369}{{\tt arXiv:1507.07369 [astro-ph]}}.

\bibitem{Bhar2016}
P.~Bhar, M.~Govender, and R.~Sharma, ``{Modeling Anisotropic Stars Obeying
  Chaplygin Equation of State},''  (2016)  1--14,
  \href{http://arxiv.org/abs/1605.01274v1}{{\tt arXiv:1605.01274v1 [gr-qc]}}.

\bibitem{Aurich2017}
R.~Aurich and S.~Lustig, ``{On the compatibility of the Chaplygin gas cosmology
  with recent cosmological data},'' \href{http://arxiv.org/abs/1704.01749}{{\tt
  arXiv:1704.01749 [astro-ph]}}.

\bibitem{Saha2016}
{S.\ Saha}, {S.\ Ghosh}, and {S.\ Gangopadhyay}, ``{Interacting Chaplygin gas
  revisited},'' \href{http://arxiv.org/abs/1701.01021}{{\tt arXiv:1701.01021
  [gen.\ physics]}}.

\bibitem{Marttens2017}
R.~F. vom Marttens, L.~Casarini, W.~Zimdahl, W.~S. Hip{\'{o}}lito-Ricaldi, and
  D.~F. Mota, ``{Does a generalized Chaplygin gas correctly describe the
  cosmological dark sector?},'' \href{http://arxiv.org/abs/1702.00651}{{\tt
  arXiv:1702.00651 [astro-ph]}}.

\bibitem{Chaplygin1904}
S.~Chaplygin, ``{On gas jets},'' {\em Sci. Ann. Imp. Univ. Mowcow} {\bf 21}
  (1904)  .

\end{thebibliography}\endgroup

\end{document}